\documentclass[aps,prl,twocolumn]{revtex4-1}  
\usepackage{amsmath,amssymb}  
\usepackage{graphicx}         
\usepackage{booktabs}
\usepackage{natbib}
\usepackage[colorlinks=true,linkcolor=blue,citecolor=blue,urlcolor=blue,bookmarksopen,breaklinks,linktoc=page]{hyperref}
\usepackage{orcidlink}

\usepackage{tabularx}
\usepackage{bm}
\usepackage{orcidlink}
\usepackage{braket}
\usepackage{subfig}
\usepackage{ragged2e} 
\usepackage{threeparttable}
\usepackage{microtype}
\usepackage[table]{xcolor}
\usepackage{tikz}
\usetikzlibrary{arrows.meta,decorations.pathreplacing,calc}

\renewcommand{\v}[1]{\boldsymbol{#1}}

\def\HIM{Helmholtz Institute Mainz, 55099 Mainz, Germany}
\def\GSI{GSI Helmholtzzentrum für Schwerionenforschung GmbH, 64291 Darmstadt, Germany}
\def\JGU{Johannes Gutenberg University, Mainz 55128, Germany}

\def\FMUV{Faculty of Mathematics, University of Vienna, Oskar-Morgenstern-Platz 1, 1090 Vienna, Austria}
\def\GPGUV{Gravitational Physics Group, University of Vienna, W\"{a}hringer Stra{\ss}e 17, 1090 Vienna, Austria}

\def\UC{Department of Physics, University of California at Berkeley, Berkeley, California 94720-7300, USA}

\begin{document}

\title{Testing Exotic Electron–Electron Interactions with the Helium Ionization-Energy Anomaly}

\author{Lei Cong$^{1,2,3}$\orcidlink{0000-0003-0002-1840}, Filip Ficek$^{4,5,*}$\orcidlink{https://orcid.org/0000-0001-5885-7064}, 
Rinat Abdullin\orcidlink{0009-0004-6992-361X},
Mikhail G.~Kozlov$^{+}$\orcidlink{0000-0002-7751-6553}
and Dmitry Budker$^{1,2,3,6}$\orcidlink{0000-0002-7356-4814}}
\address{
$^{1}$\HIM\\
$^{2}$\GSI\\
$^{3}$\JGU\\
$^{4}$\FMUV\\
$^{5}$\GPGUV\\
$^{6}$\UC\\
* filip.ficek@univie.ac.at; 
$^+$ mihailgkozlov@gmail.com
 }

\begin{abstract}
Precision atomic spectroscopy provides a sensitive probe of physics beyond the Standard Model. A recently reported $9\sigma$ theory–experiment discrepancy in the ionization energy of metastable helium has motivated the hypothesis of a new boson mediating exotic electron–electron interactions. Using a model-independent sign-consistency analysis of the induced energy shifts, we show that the sign requirement alone excludes vector–vector and pseudoscalar–pseudoscalar interactions as possible explanations of the anomaly. 
Incorporating existing constraints together with independent limits obtained here further excludes axial-vector scenarios.
Within the single-boson framework considered in this work, only a narrowly constrained scalar-mediated interaction remains viable.
The remaining parameter space could be probed, for example, by modest improvements in the determination of the electron gyromagnetic ratio.

\end{abstract}

\maketitle

\textit{Introduction.}--- The pursuit of physics beyond the Standard Model (BSM) is motivated by several
well-established open questions in fundamental physics, including the origin
of the matter–antimatter asymmetry, the nature of dark matter and dark energy,
and the hierarchy problem \cite{dine_origin_2003,feng_dark_2010,graham_cosmological_2015,navas_review_2024}.
Among the
various experimental approaches proposed to probe BSM physics
\cite{safronova_search_2018,cong_spin-dependent_2025,cosmicwispers_wg4_2025}, precision atomic spectroscopy has emerged as a powerful
low-energy probe, linking high-accuracy measurements to possible new
interactions beyond the SM.
In particular, simple few-body atomic systems such as hydrogen
\cite{karshenboim_precision_2010}, positronium \cite{adkins_precision_2022},
and helium \cite{delaunay_probing_2017,ficek_constraints_2017} provide
exceptionally clean testbeds, where discrepancies between theory and experiment
can offer sensitive probes of previously unknown forces.


Helium spectroscopy has long served as a cornerstone of experimental
\cite{schuessler_hyperfine_1969,sun_precision_2020,clausen_metrology_2025,clausen_ionization_2025} and
theoretical \cite{born_atomic_1989,bethe_quantum_2013,pachucki_testing_2017,patkovs_complete_2021,drake_high_2023}
quantum physics, enabling stringent tests of quantum electrodynamics
\cite{rengelink_precision_2018} and the determination of fundamental constants
\cite{sun_precision_2020}. A long-standing agreement between theory and
experiment has recently been challenged by the report of a $9\sigma$
discrepancy in the ionization energy of the metastable $2^{3}S_{1}$ state of
$^{4}$He \cite{clausen_metrology_2025,patkovs_complete_2021}, which has now also
been observed in $^{3}$He \cite{clausen_ionization_2025}. The persistence of the
anomaly across the two isotopes has therefore been discussed as a possible
indication of new physics \cite{bondy_precision_2025}.

\begin{figure}[t]
    \centering
    \includegraphics[width=0.95\linewidth]{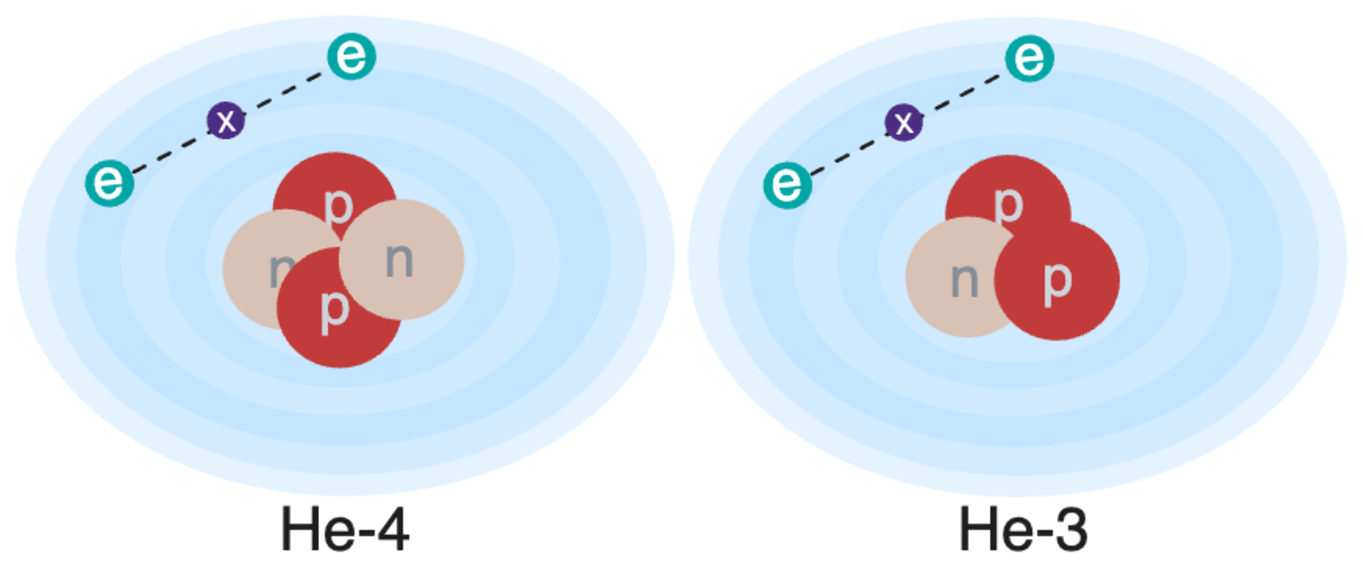}
  \caption{Schematic illustration of the assumed exotic interaction, mediated by a new boson ($X$), between electrons within $^{4}$He and $^{3}$He.
Only electron--electron exotic interactions are considered, as they generate identical leading-order energy shifts in both isotopes despite their different nuclear structures. The shaded regions indicate schematic electron probability distributions.
}
    \label{fig:schematic}
\end{figure}

Building on this observation, we note that the comparable magnitude of the
discrepancies in $^{4}$He and $^{3}$He strongly disfavors explanations rooted in
nuclear structure, since the helion and the alpha particle differ substantially
in their composition and spin. In particular, finite-size nuclear effects have
been excluded as the origin of the anomaly in
Ref.~\cite{steinebach_spectroscopy_2026}.
This consideration instead points to an
interaction between the electrons in helium (see Fig.\,\ref{fig:schematic}), affirming helium
spectroscopy as an exceptionally sensitive low-energy
probe of new particles, operating in a regime complementary to that explored by
high-energy colliders.

From the perspective of new-boson searches, the emergence of a discrepancy at the $9\sigma$ level marks a qualitative turning point. The anomaly is now sufficiently large that the sign of any proposed new-physics contribution becomes a decisive consistency criterion, rather than a sub-leading detail. This enables a fundamentally different approach to testing exotic interactions, going beyond the incremental tightening of existing bounds~\cite{cong_spin-dependent_2025}.

In this Letter, we take the reported helium discrepancy at face value and perform
a stringent test of the single-boson hypothesis. We establish a sign-consistency criterion that becomes decisive at the 9$\sigma$ level. We compute the
exotic-interaction–induced corrections to helium ionization-energy for four
generic coupling structures—$g_A g_A$, $g_s g_s$, $g_V g_V$, and $g_p g_p$ (where the superscripts indicate axial vector, scalar, vector, and pseudoscalar couplings). 
These couplings arise from the exchange of exotic spin-0 and spin-1 bosons,
such as axion-like particles and $Z'$ bosons \cite{safronova_search_2018,cong_spin-dependent_2025},
between two electrons.
We examine whether the \textit{signs} and magnitudes of the couplings can consistently account for the
observed deviation. 
We find that pseudoscalar and vector interactions are immediately excluded by
sign inconsistency. A consistent reinterpretation of existing constraints,
together with the improved limits obtained here, then rules out axial-vector
scenarios, leaving a scalar coupling as the only partially viable possibility
compatible with all current data.

\textit{Theory.}--- 
The theory of exotic interactions mediated by a ``new'' boson has been developed over the past four decades~\cite{moody_new_1984,dobrescu_spin-dependent_2006,fadeev_revisiting_2019}, as summarized in Ref.\,\cite{cong_spin-dependent_2025}, where the sixteen possible exotic potentials are shown to be grouped into nine distinct coupling types.
In this work we consider four classes: axial-vector--axial-vector (AA), pseudoscalar--pseudoscalar (pp), vector--vector (VV), and scalar--scalar (ss). The relevant potentials are the spin-dependent
$V|_{AA}$;
$V|_{pp}$;
$V|_{VV}$;
and spin-independent $V|_{ss}$. Owing to their length and partial redundancy, only $V|_{AA}$ and $V|_{ss}$ are shown in the main text; the others appear in the Supplemental Material.
\begin{widetext}
\begin{equation}
\label{gaga_V3}
\begin{gathered}
V_{AA}= \underbrace{-g_A^eg_A^e \frac{\hbar c}{4\pi}\boldsymbol{\sigma}_e\cdot\v\sigma_e^{\,\prime}\frac{1}{r}\,e^{-{r}/{\lambda}} }_{{V_2}|_{AA}} \\
 \underbrace{-  {g_A^eg_A^e } \frac{\hbar c} {4 \pi} \lambda^2  \left[ \v{\sigma}_e \cdot \v\sigma_e^{\,\prime}\left[ \frac{1}{r^3} + \frac{1}{\lambda r^2} + \frac{4 \pi}{3} \delta(\v{r}) \right] -  \left( \v{\sigma}_e \cdot \hat{\v{r}} \right) \left( \v\sigma_e^{\,\prime}\cdot \hat{\v{r}} \right)  \left( \frac{3}{r^3} + \frac{3}{\lambda r^2} + \frac{1}{\lambda^2 r} \right)  \right] {e^{-r/\lambda}}}_{{V_3}|_{AA}}\,,
\end{gathered}
\end{equation}


\end{widetext}
\begin{equation}
\label{gsgs_V1}
V_{ss} = \underbrace{- g^e_s g^e_s \frac{\hbar c}{4\pi}\frac{e^{-{r}/{\lambda}}}{r}}_{V_1|_{ss}} \, ,
\end{equation}
where $\hbar$ is the reduced Planck constant, $c$ is the speed of light in vacuum, $\boldsymbol{\sigma}_{e}$ and $\v\sigma_e^{\,\prime}$ are vectors of Pauli matrices representing the spins $\boldsymbol{s}_i=\hbar \boldsymbol{\sigma}_i/2$ of the two electrons, 
$m_{e}$ is the electron mass, 
$M$ is the new boson mass, which is inversely proportional to the force range $\lambda$, $M=\hbar/(c\lambda)$, and $r$ is the distance between the two electrons. 



\begin{table*}[htbp]
\centering
\begin{threeparttable}
\caption{\textbf{Theoretical and experimental inputs for calculating $\Delta E$.}
The relevant ionization and transition energies are listed; ionization energies are given relative to the ionization threshold, such that bound-state energies are negative.
$\mu = \text{Expt} - \text{Theory}$, 
$\sigma = \sqrt{\sigma_{\mathrm{th}}^{2} + \sigma_{\mathrm{expt}}^{2}}$.
The quantity $L$ is derived from the integral equation given below Eq.~\eqref{eq:deltaE}.
For the new-boson case, $\Delta E = \mu \pm L$.
Note that, for comparison with theory, we combine statistical and systematic uncertainties of the experimental results for ionization energy in quadrature.}
\label{tab:DeltaE}

\renewcommand{\arraystretch}{1.3}
\small

\begin{tabular*}{\textwidth}{@{\extracolsep{\fill}} l c c c c c}
\hline\hline
Parameter & Theory (kHz) & Exp (kHz) & $\mu$ (kHz) & $\sigma$ (kHz) & $\Delta E$ (kHz)\\
\hline
 ${^4}He$ 2$^3S_1$ &-1\,152\,842\,742\,231 (52)\,\cite{patkovs_complete_2021} &-1\,152\,842\,742\,708.2 (60)\,\cite{clausen_metrology_2025}& -477 & 52  & -477 $\pm$ 102\\
\hline
${^3}He$ 2$^3S_1$&-1\,152\,788\,844\,133\,(52)\,\cite{pachucki_testing_2017,patkovs_complete_2021,pachucki_theory_2015}&-1\,152\,788\,844\,615.4\,(81)\,\cite{clausen_ionization_2025}& -482 &53 &-482 $\pm$103\\
\hline
\hline
 2$^3S_1$-2$^3P_0$  &  276\,764\,094\,677\,(54)\,\cite{pachucki_fine_2010,pachucki_testing_2017,patkovs_complete_2021}\tnote{*} &276\,764\,094\,712.45\,(0.86)\,\cite{wen_postselection_2025} & -35 &54 & 125\\
\hline\hline
\end{tabular*}

\begin{tablenotes}
\footnotesize
\item[*] See the Supplemental Material for the calculation of the theoretical values used.
\end{tablenotes}

\end{threeparttable}
\end{table*}

\begin{figure}[!htbp]
    \centering
    \includegraphics[width=1\linewidth]{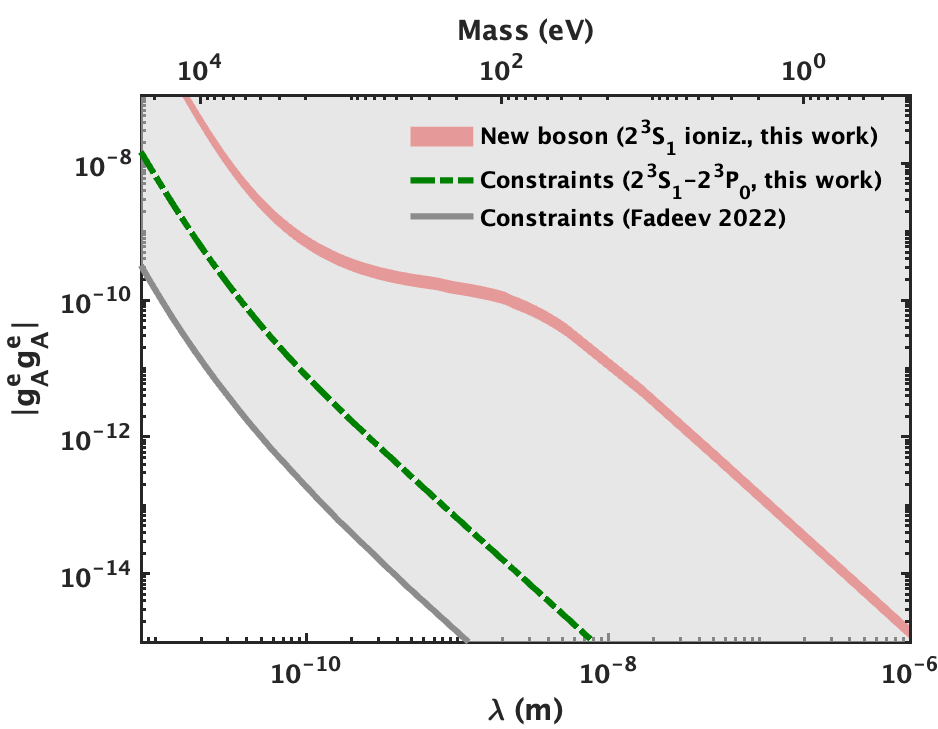}
  \caption{
Axial-vector new-boson coupling product $g_A^e g_A^e$ inferred from the helium discrepancy (pink band), shown together with existing constraints (gray curves) and the new constraints obtained in this work (green dash-dotted curves) as a function of the interaction range $\lambda$ (bottom axis). The corresponding boson mass $M$ is shown on the top axis.
For an axial-vector boson, the coupling implied by the helium discrepancy is excluded by our new constraints, in combination with previous limits from Ref.~[6].
}
    \label{fig:gAgA}
\end{figure}

\begin{figure}
    \centering
    \includegraphics[width=1\linewidth]{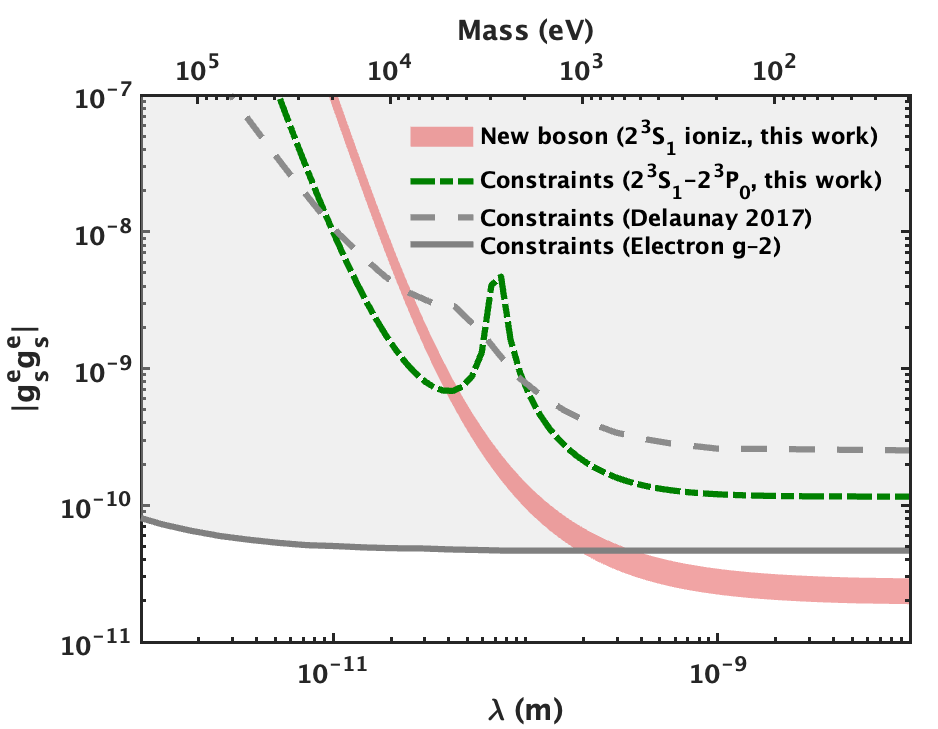}
    \caption{
   Scalar new-boson coupling product $g_s g_s$ inferred from the helium discrepancy (pink band), shown together with the new constraints obtained in this work (green dash-dotted curve) and existing constraints from Ref.~\cite{delaunay_probing_2017} (gray curves) as a function of the interaction range $\lambda$. The corresponding boson mass $M$ is shown on the top axis. In this case, the coupling implied by the helium discrepancy remains possible for $M < 800\,\mathrm{eV}$, while the higher-mass range is excluded by existing and new constraints.    }
    \label{fig:gsgs}
\end{figure}

\textit{Method.}--- 
To extract the coupling strength of a hypothetical new boson from the theory--experiment discrepancy observed in the ionization energy
of the $2^3S_1$ state of helium, we interpret the discrepancy as arising
entirely from an exotic interaction.
The corresponding energy shift is obtained by evaluating the matrix element
of the exotic potential using the electronic wavefunction of this state,
\begin{equation}\label{DeltaE1}
E^{\rm exotic}
= \bra{\Psi_{2^3S_1}} V_i \ket{\Psi_{2^3S_1}} \, .
\end{equation}
The resulting shift can be written in the form
$E^{\rm exotic} = g^e g^e\, C(\lambda)$, where the overall magnitude is fixed
by the difference between the most recent experimental and theoretical
determinations,
\begin{align}\label{eq:deltaE}
\Delta E
= E_{\mathrm{exper}} - E_{\mathrm{theor}}
= \mu \pm L \, ,
\end{align}
where $\mu$ denotes the central value of the discrepancy and $L$ represents
the combined experimental and theoretical uncertainty at the $95\%$
confidence level (CL), defined through
$\int_{-L}^{L}
\frac{1}{\sqrt{2\pi}\sigma}\,
e^{-x^{2}/(2\sigma^{2})}\, dx
= 0.95$.
Under this interpretation, the corresponding coupling strength is given by
$g^e g^e = {\Delta E}/{C(\lambda)}$.
In the present case, we extract the coupling strength from the upper and lower bounds, $\mu \pm L$  (see Table~\ref{tab:DeltaE}), and display it as a band corresponding to this, see Fig.\,\ref{fig:gAgA} and \ref{fig:gsgs}.
On the other hand, recent improvements in measurements of the hyperfine
structure of helium \cite{zheng_laser_2017} and of the fine-structure intervals
\cite{zheng_measurement_2017,wen_postselection_2025} enable a precise
determination of the $2{}^3S_1\!-\!2{}^3P_0$ transition (see Table~\ref{tab:DeltaE}).
In this case, theoretical predictions and experimental results are in
excellent agreement, allowing substantially stronger constraints to be set
than in earlier studies \cite{ficek_constraints_2017} (See more in the Supplement Materials).



\textit{Results.}--- 
We first show that sign consistency alone already excludes
two of the four generic coupling structures, independent of any quantitative
constraints. In particular, pseudoscalar- and vector-mediated exotic
electron–electron interactions are incompatible with the observed helium
ionization-energy discrepancy.

This conclusion follows from a simple but decisive observation. For $e$–$e$
exotic interactions, the relevant couplings enter only through their squares,
${g_A^e}^2$, ${g_p^e}^2$, ${g_s^e}^2$, and ${g_V^e}^2$, which are strictly
positive. As a result, each exotic potential contributes with a fixed sign that
is uniquely determined by its operator structure. Since the measured and
theoretical energies satisfy $E^{\mathrm{exper}} =
E^{\mathrm{theory}} + E^{\mathrm{exotic}}$, the helium data in
Table~\ref{tab:DeltaE} require $E^{\mathrm{exotic}}$ to be negative. This sign
requirement immediately rules out pseudoscalar interactions and vector-type
potentials of the ${V_1+V_2+V_3}|_{VV}$ form, whose predicted shifts have the
opposite sign.

In earlier analyses of exotic-interaction constraints, such as
Ref.\,\cite{ficek_constraints_2017}, theory–experiment differences were much
smaller and dominated by combined uncertainties spanning both signs, rendering
the sign of the exotic contribution effectively irrelevant. Here, by contrast,
the much larger $9\sigma$ discrepancy elevates sign consistency to a decisive
criterion.

Having eliminated pseudoscalar and vector scenarios by sign consistency alone,
we next examine whether the remaining axial-vector and scalar interactions are
compatible with existing experimental constraints.

Our results are summarized in
Figs.\,\ref{fig:gAgA} and \ref{fig:gsgs}. In both figures, the pink bands denote
the coupling strengths required to account for the helium discrepancy, as
extracted from $^{4}$He data. The corresponding coupling band inferred from $^{3}$He is shifted by approximately $1\%$ relative to the $^{4}$He central value, while its width is nearly identical to that of the $^{4}$He band, differing below the $10^{-3}$ level; it lies slightly above the $^{4}$He curve and is therefore not shown for clarity.
The dash-dotted green curves indicate the new
constraints obtained in this work, while the gray curves represent existing
direct or indirect bounds.

Figure~\ref{fig:gAgA} demonstrates that an axial-vector boson
explanation of the helium discrepancy is incompatible with the constraints based on combined exotic potentials. Because a physical axial-vector mediator generically induces all
allowed axial–axial interaction terms, the relevant limits must be evaluated
for the full combined potential $V_2+V_3|_{AA}$ \cite{cong_spin-dependent_2025,cong_improved_2025,kang_exotic-interaction_2025,fadeev_pseudovector_2022} rather than for individual
contributions (see more in the Supplement Materials). Accordingly, Fig.~\ref{fig:gAgA} compares the coupling
required to explain the helium discrepancy with constraints derived from the
combined $V_2+V_3|_{AA}$ interaction. 
The new bounds from the $2^3S_1$--$2^3P_0$ transition (see Table\,\ref{tab:DeltaE})
obtained in this work and the previously established limits \cite{fadeev_pseudovector_2022}
both exclude this scenario entirely.

We now turn to the spin-independent scalar interaction. In this case, the induced energy shift has the same sign as the observed discrepancy and therefore survives the sign-consistency test. 
Using a wavefunction for the $2^{3}S_{1}$ state from Refs.\,\cite{bethe_quantum_2013,ficek_constraints_2017}, we find that, for a scalar mediator with $M < 800\,\mathrm{eV}$, the coupling strength required to reproduce the reported discrepancy lies below existing bounds, leaving room for a possible explanation of the discrepancy.

In more detail, Fig.\,\ref{fig:gsgs} shows that a spin-independent scalar boson remains allowed
when compared with previous constraints (gray dotted line) derived from the
$2\,{}^{1}\!S_{0}\!-\!2\,{}^{3}\!S_{1}$ and
$2\,{}^{3}\!P_{0}\!-\!3\,{}^{3}\!D_{1}$ transitions in helium
\cite{delaunay_probing_2017}, permitting a mass range of
$M < 8000\,\mathrm{eV}$. However, this remaining parameter space is reduced to
$M < 5000\,\mathrm{eV}$ by our newly obtained constraints (dash-dotted green line)
from the $2^{3}S_{1}$--$2^{3}P_{0}$ helium transition. To further narrow the
possible mass window for a new boson, we also show the electron $g\!-\!2$
constraint~\cite{delaunay_probing_2017} for scalar bosons in
Fig.\,\ref{fig:gsgs}. This bound is stronger and further excludes a scalar-boson explanation of the
helium discrepancy, leaving it possible only for $M < 800\,\mathrm{eV}$.


Comparisons with state-of-the-art helium wave-function calculations \cite{AbdullinKozlov_inprep} show that
improvements to the electronic wavefunction modify the relevant matrix elements
by factors of order unity at most (see Supplemental Material), which is insufficient to qualitatively alter
the conclusions presented above.
We note, however, that the electron $g\!-\!2$ constrain new physics \cite{leveille_second-order_1978,LINDNER20181} in a
conceptually different manner from fifth-force searches. In helium spectroscopy,
a new boson is treated as being exchanged between two electrons, giving rise to an
effective exotic potential \cite{cong_spin-dependent_2025}, whereas in the $g\!-\!2$ measurement it enters
through virtual loop corrections to the electron--photon vertex
\cite{na64_collaboration_constraints_2021,frugiuele_current_2019,
delaunay_probing_2017}. As a result, the corresponding $g\!-\!2$ bounds are subject to the
assumed ultraviolet completion of the interaction and are therefore not directly
equivalent to constraints from fifth-force experiments.
Further direct constraints on $g_s g_s$, for example, from new measurements of helium transitions, are desirable.

\textit{Conclusion}--- We have examined whether exotic spin-dependent or spin-independent interactions
can account for the reported $9\sigma$ discrepancy in the ionization energy of
the metastable $2^{3}S_{1}$ states of $^{3}$He and $^{4}$He. A sign analysis of
the induced energy shifts immediately excludes pseudoscalar–pseudoscalar
($g_p g_p$) and vector–vector ($g_V g_V$) couplings, as they generate shifts of
the wrong sign. Among the remaining possibilities, an axial-vector interaction
($g_A g_A$) is excluded once the full $V_2+V_3|_{AA}$ structure and existing
constraints are consistently taken into account. A scalar interaction
($g_s g_s$) remains the only partially viable scenario: our new constraints
from the $2^{3}S_{1}$--$2^{3}P_{0}$ helium transition rule it out down to
$M < 5000\,\mathrm{eV}$, while indirect electron $g\!-\!2$ limits further
disfavor masses above $M \gtrsim 800\,\mathrm{eV}$.
Overall, the observed discrepancy is strongly disfavored as a signal of a new
single boson, except possibly for a light scalar mediator with
$M < 800\,\mathrm{eV}$ within the single-boson, single-coupling framework
considered here.

Even this scalar window, however, is already severely constrained, underscoring
that any viable new-physics explanation must be carefully disentangled from
potentially missing theoretical contributions.
One intriguing clue is that, while the measured
$^{3}S_{1}$–$^{3}P_{J}$ transition frequencies agree remarkably well with
state-of-the-art calculations~\cite{wen_postselection_2025,pastor_absolute_2004,patkovs_complete_2021},
the calculated binding energies of the individual
$(1s)(2s)\,{}^{3}S_{1}$ and $(1s)(2p)\,{}^{3}P_{J}$ levels each deviate at the
$\sim0.5\,\mathrm{MHz}$ level~\cite{clausen_precision_2025}. The apparent agreement
in the transition frequencies may therefore arise from a coincidental
cancellation, highlighting the importance of revisiting higher-order QED
contributions, including the $\alpha^{7}$ terms.

Finally, our approach is not limited to systems
exhibiting anomalies but is applicable to a wide class of
precision platforms, including muonic atoms, atomic clocks, and related
systems.

\textit{Acknowledgements}--- 
The authors acknowledge helpful discussions with Frederick Merkt, Gordon Drake, Yevgeny Stadnik, 
Vladimir Yerokhin,
Shui-Ming Hu,
and Danial Saeed.
This work has been supported by the Cluster of Excellence “Precision Physics,
Fundamental Interactions, and Structure of Matter” (PRISMA++ EXC 2118/2) funded
by the German Research Foundation (DFG) within the German Excellence Strategy
(Project ID 390831469) and by the COST Action within the project COSMIC WISPers (Grant No. CA21106).
F. F. acknowledges the support of the Austrian Science Fund (FWF) via Project No. PAT 9429324 (DOI:10.55776/PAT9429324).
R.\,A.\ and M.\,K.\ were supported by the Foundation for the Advancement of Theoretical Physics and Mathematics ``BASIS''.

\bibliographystyle{apsrev4-1} 
\bibliography{MyLibrary}

\end{document}